\documentclass[aps,prd,nofootinbib,superscriptaddress]{revtex4}
\usepackage[sumlimits]{amsmath} 
\usepackage{epsfig}
\usepackage{amssymb}
\usepackage{graphicx}
\usepackage{pstricks}
\usepackage{color}
\usepackage{bbold}
\usepackage{hyperref}

\begin{document}

%
%%%%%%%%%%%%%%%%%%%%%%%%%%%%%%%%%%%%%%%%%%%%%%%%%
%%%%%%%%%%%%%%%%   TITLE   %%%%%%%%%%%%%%%%%%%%%%
%%%%%%%%%%%%%%%%%%%%%%%%%%%%%%%%%%%%%%%%%%%%%%%%%
%
\title{Cosmological Boundary Flux Parameter}

\author{Rafael Hern\'{a}ndez-Jim\'{e}nez}
\email{rafaelhernandezjmz@gmail.com}
\author{Claudia Moreno} 
\email{claudia.moreno@cucei.udg.mx}
\affiliation{Departamento de F\'isica,
Centro Universitario de Ciencias Exactas y Ingenier\'ias, Universidad de Guadalajara
Av. Revoluci\'on 1500, Colonia Ol\'impica C.P. 44430, Guadalajara, Jalisco, M\'exico}
\author{Mauricio Bellini} 
\email{mbellini@mdp.edu.ar}
\affiliation{Departamento de F\'{i}sica, Facultad de Ciencias Exactas y Naturales, Universidad Nacional de Mar del Plata, Funes 3350, C.P. 7600, Mar del Plata, Argentina \\
Instituto de Investigaciones F\'{i}sicas de Mar del Plata (IFIMAR), Consejo Nacional de Investigaciones Cient\'{i}ficas y T\'{e}cnicas (CONICET), Mar del Plata, Argentina}

\author{C. Ortiz} 
\email{ortizgca@fisica.uaz.edu.mx}
\affiliation{Unidad Acad\'{e}mica de F\'{i}sica, Universidad Aut\'{o}noma de Zacatecas, \\
Calzada Solidaridad esquina con Paseo a la Bufa S/N C.P. 98060, Zacatecas, M\'{e}xico}
%
%%%%%%%%%%%%%%%%
%%%%%%   DATE   %%%%%
%%%%%%%%%%%%%%%%
%
\date{\today}

%
%%%%%%%%%%%%%%%%%%%%%%%%%%%%%%%%%%%%%%%%%%%%%%%%%%
%%%%%%%%%%%%%%%%%%   ABSTRACT   %%%%%%%%%%%%%%%%%%
%%%%%%%%%%%%%%%%%%%%%%%%%%%%%%%%%%%%%%%%%%%%%%%%%%
%
\bigskip
\begin{abstract}
The {\it{Cosmological Boundary Flux Parameter}} is a novel proposal that attempts to explain the origin of the cosmological parameter $\Lambda$ purely by geometric nature. Then we implement this new approach to a flat FLRW universe along with a barotropic fluid. We present an ansatz in which $\Lambda$ is straightforwardly coupled to the matter sector; therefore, only one additional parameter was introduced: $\lambda$. Also, through a statistical analysis, using late-time data of observational Hubble and type Ia Supernovae, we computed the joint best-fit value of the free parameters by means of the affine-invariant MCMC. We want to emphasise that the joint analysis produces a smaller $H_{0}^{\rm CBFP}=69.80\rm\,\, Km \,s^{-1}\,Mpc^{-1}$ in contrast to the flat $\Lambda$CDM result $H_{0}^{\Lambda\rm CDM}=70.53\rm\,\, Km \,s^{-1}\,Mpc^{-1}$. The work presented here seeks to contribute to the discussion of the possible explanation for the cosmos' acceleration, together with tackling other important questions in modern cosmology.
\end{abstract}

% \pacs{98.80.Cq, 11.10.Wx, 14.80.Bn, 14.80.Va}
% \pacs{98.80.-k, 04.50.Kd, 95.36.+x, 98.80.Es}

\maketitle

%
%%%%%%%%%%%%%%%%%%%%%%%%%%%%%%%%%%%%%%%%%%%%%%%%%%%%%%%%%%%%
%%%%%%%%%%%%%%%%%%% INTRODUCTION %%%%%%%%%%%%%%%%%%%%%%%%%%%
%%%%%%%%%%%%%%%%%%%%%%%%%%%%%%%%%%%%%%%%%%%%%%%%%%%%%%%%%%%%
%
\section{Introduction}
%
%%%%%%%% UN POCO DE COSMO, MODELO LCDM,  %%%%%%%%%%%%%%%%%%%%%%%% 
%%%%%%%% ENERGÍA OBSCURA Y TENSIÓN DE H0 %%%%%%%%%%%%%%%%%%%%%%%%
Certainly General Relativity (GR), described by the Einstein equations, is nowadays the most accurate description of several gravitational phenomena. Indeed, GR sets the framework in which cosmology lies. The standard cosmological model, in addition to radiation and barionic matter, incorporates the so-called Cold Dark Matter (CDM) component and the Cosmological Constant $\Lambda$. Both elements constitute the dark sector. The former may explain the formation of a large structure \cite{1982Natur.299...37B,1982PhRvL..48.1636B,1982PhRvL..48..223P,1982ApJ...258..415P,1983ApJ...274..443B,1984MNRAS.211..277D}, along with other astrophysical phenomena such as the flatness on the galaxy rotation curve \cite{1970ApJ...160..811F,1970ApJ...159..379R}; and CDM only interacts gravitationally with the rest of the known particles. The latter was proposed to explain the current epoch of accelerated expansion of our universe, discovery established in 1998 independently from the High-redshift Supernova Search team and the Supernova Cosmology Team, led by Adam Riess \cite{SupernovaSearchTeam:1998fmf} and Saul Perlmutter \cite{SupernovaCosmologyProject:1998vns} respectively, collected distances for 51 Supernovae Type Ia (SNe Ia). Thus, the aforementioned elements yield the $\Lambda$CDM paradigm. Indeed, this model remains the simplest candidate that yields a good fit to a large collection of cosmological data; yet, areas of phenomenology and ignorance arise. Nevertheless, since 1988-1989 Steven Weinberg has already described how the cosmological constant presents issues from both perspectives: modern theories of elementary particles and astronomical observations~\cite{Weinberg:1988cp}. 

A plethora of dark energy proposals have been put forward. From scalar fields~\cite{Caldwell:1999ew,Caldwell:2003vq,Nojiri:2005sx,Feng:2006ya,Linder:2007wa,Setare:2008sf,Tsujikawa:2013fta,Chiba:2012cb,Linde:2015uga,Linder:2015qxa,Durrive:2018quo,Bag:2017vjp,Leon:2018lnd,Garcia-Garcia:2018hlc,Alestas:2020mvb}; as well as fluids with variable equation of state~\cite{Carturan:2002si,Cardone:2005ut,Nojiri:2006zh,Brevik:2007jt,Linder:2008ya,Duan:2011jj,Bini:2013ods,Barrera-Hinojosa:2019yyh}; and modified gravity~\cite{Lobo:2008sg,Clifton:2011jh,Dimitrijevic:2012kb,Brax:2015cla,Joyce:2016vqv,Jaime:2018ftn,Slosar:2019flp}. More recently have been presented alternative proposals; one is the cosmological diffusion models in Unimodular Gravity \cite{Corral:2020lxt,LinaresCedeno:2020uxx}, which is a framework where $\Lambda$ is not a term introduced by hand, as historically Einstein presented it \cite{Einstein:1917ce}, but it appears straightforwardly as an integration constant when considering the Einstein–Hilbert action with volume-preserving diffeomorphisms.

% On the other hand, due to current highly improved observational surveys, conflicts among key cosmological parameters of the model have emerged. One of these discrepancies is the current value of the Hubble parameter $H_{0}$. When the Hubble parameter is measured on local astrophysical scales by scaling methods from cosmic ladders, the values obtained for $H_0$ are different from those reported by Planck, which depend on the standard $\Lambda$CDM model and uses the Cosmic Microwave Background (CMB) photons. This difference is known as the Hubble tension~\cite{Addison:2017fdm, DES:2018rjw,Wong:2019kwg} and has a significance of $5\sigma$ if one considers the CMB observations~\cite{Planck:2018nkj}, a value of $H_{0}^{\rm CMB}=67.66\pm 0.42 \rm\,\, Km \,s^{-1}\,Mpc^{-1}$ (Planck + BAO), and the local cosmic distance ladder~\cite{Riess:2020fzl}, where the latest Cepheids with the Hubble Space Telescope yield $H_{0}^{\rm Cepheids}=73.2\pm 1.3\rm\,\, Km \,s^{-1}\,Mpc^{-1}$. In the absence of unknown systematic errors, this discrepancy could suggest the existence of physics beyond the standard cosmological paradigm. This scenario has led cosmologists to propose and study new cosmological models, mainly, but not limited to those that extend the $\Lambda$CDM model at early or late times~\cite{Abdalla:2022}. A titanic summary of this matter can be found in~\cite{DiValentino:2021izs}. 

%%%%%%%%%% TÉRMINO DE BORDE %%%%%%%%%%%%%%%%%%%%%%%%%%%%%%%%%%%%%%%%%
We modestly present a novel scheme that attempts to explain the origin of the cosmological parameter $\Lambda$ purely by geometric nature. To begin with the Lagrangian formulation of GR from the Einstein-Hilbert (EH) action, formed with the only independent scalar constructed from the metric, which is no higher than second order in its derivatives, the Ricci scalar $R = g^{\mu\nu}R_{\mu\nu}$ times the square root of the negative determinant of the metric tensor $\sqrt{-g}$. Then, the equations of motion should arise from the variation of the action with respect to the metric $g^{\mu\nu}$. Indeed, the variation of the Ricci tensor $\delta R_{\mu\nu}$ yields the covariant divergence of a vector which by Stokes' theorem is equal to a boundary contribution at infinity which we can set to zero by making the variation vanish at infinity. However, when the underlying spacetime manifold has a boundary $\partial V$, aforementioned procedure leads to a cumbersome assumption. To solve this inelegant issue, Hawking-Gibbons-York (HGY) proposed adding a counterterm in the EH action, which relates the boundary constraint and extrinsic curvature \cite{Gibbons:1976ue, York:1972sj}, to cancel such input. However, by eliminating this extremum, any physical phenomena at the border are excluded, provided that they are indeed considered irrelevant. Instead of dropping the boundary expression, an alternative proposal is to take it into account as a physical source of geometric nature~\cite{Ridao:2015oba, Ridao:2014kaa}. We name {\it{Cosmological Boundary Flux Parameter}} (CBFP) this novel approach. The work presented here seeks to contribute to the discussion of the possible explanation for the cosmos' acceleration, together with tackling other important questions in modern cosmology using late-time observations.

This paper is organised as follows. In section~\ref{Origin_CP} we briefly introduce the cosmological parameter as a result of boundary conditions due to a close spacetime manifold. In section~\ref{BCP}, we present a practical example, a flat Friedmann-Lema\^{i}tre-Robertson-Walker (FLRW) universe, along with a barotropic fluid. We take a particular ansatz for $\Lambda$ in order to analytically solve the set of specific differential equations. Then, section~\ref{cosmo-constraints} shows a description of the methods and the data (SNe Ia and observational Hubble data) used to constrain the CBFP model. Subsequently, section~\ref{analysis_results} displays the best statistical estimate of the constrained parameters due to different astrophysical observations. Finally, in section~\ref{conclusions} we will give the conclusion and outlook of this work.
%
%%%%%%%%%%%%%%%%%%%%%%%%%%%%%%%%%%%%%%%%%%%%%%%%%%%%%%%%%%%%
%%%%%%%% ORIGIN OF THE COSMOLOGICAL PARAMETER %%%%%%%%%%%%%%
%%%%%%%%%%%%%%%%%%%%%%%%%%%%%%%%%%%%%%%%%%%%%%%%%%%%%%%%%%%%
%
\section{Origin of the cosmological parameter}\label{Origin_CP}
We begin with the EH action described gravitation and matter in our universe, and it is represented by:
\begin{equation}\label{general_action}
\mathcal{S}=\int d^{4}x\sqrt{-g}\left[\frac{R}{2\kappa}+\mathcal{L}_{m} \right]\,,
\end{equation}
where $\kappa = 8\pi G/c^{4}$, $g$ is the determinant of the covariant background tensor metric $g_{\mu\nu}$, $R=g^{\mu\nu}R_{\mu\nu}$ and $R^{\alpha}_{\,\mu\nu\alpha}=R_{\mu\nu}$ are the scalar curvature and the Ricci curvature tensor, respectively. They are derived from the curvature tensor $R^{\alpha}_{\,\beta\gamma\delta}= \Gamma^{\alpha}_{\,\beta\delta\,,\gamma} - \Gamma^{\alpha}_{\,\beta\gamma\,,\delta} + \Gamma^{\epsilon}_{\,\beta\delta}\Gamma^{\alpha}_{\,\epsilon\gamma} - \Gamma^{\epsilon}_{\,\beta\gamma}\Gamma^{\alpha}_{\,\epsilon\delta}$, where Christoffel symbols are written in terms of the metric tensor and its partial derivatives $\Gamma^{\sigma}_{\alpha\beta} = \left(g_{\gamma\beta\,,\alpha}+g_{\gamma\alpha\,,\beta}-g_{\alpha\beta\,,\gamma}\right)g^{\sigma\gamma}/2$. The Greek indices run from 0 to 3, additionally if latin indices m, n, etc. appear, they go from 1 to 3. Finally, $\mathcal{L}_{m}$ is an arbitrary Lagrangian density that describes matter. First of all, let us recall that the standard procedure to obtain the Einstein field equations requires a variation $\mathcal{S}$ with respect to the metric tensor, and in fact, let us consider variations with respect to the inverse metric $g^{\mu\nu}$. Accordingly, we have the variation of the action matter
\begin{equation}
\delta\left[\sqrt{-g}\mathcal{L}_{m}(g^{\mu\nu},g^{\mu\nu}_{\,\,\,,\lambda})\right] = - \frac{\sqrt{-g}}{2}\delta g^{\mu\nu} T_{\mu\nu} \,,  
\end{equation}
here, we have used the generic definition of the stress-energy tensor:
\begin{equation}\label{stress-energy_tensor1}
T_{\mu\nu} = g_{\mu\nu}\mathcal{L}_{m} - 2\frac{\delta\mathcal{L}_{m}}{\delta g^{\mu\nu}}  \,.  
\end{equation}
Now, we consider the gravitational action. Its variation is
\begin{equation}\label{vary_lagrange}
\delta\left[\sqrt{-g}R\right] = \sqrt{-g}\left[\delta g^{\alpha\beta}\, G_{\alpha\beta}+g^{\alpha\beta}\delta R_{\alpha\beta}\right] \,,   
\end{equation}
where $G_{\alpha\beta}=R_{\alpha\beta} - g_{\alpha\beta}R/2$ is the Einstein tensor and 
\begin{equation}\label{boundary-term1}
g^{\alpha\beta}\delta R_{\alpha\beta} = \nabla_{\mu}\left[g^{\alpha\beta}(\delta\Gamma^{\mu}_{\alpha\beta}) - g^{\alpha\mu}(\delta\Gamma^{\gamma}_{\alpha\gamma})\right]  \,,    
\end{equation}
where $\delta\Gamma^{\mu}_{\alpha\beta}$ is an arbitrary variation of the connection, introduced by replacing $\Gamma^{\mu}_{\alpha\beta}\rightarrow \Gamma^{\mu}_{\alpha\beta}+\delta\Gamma^{\mu}_{\alpha\beta}$. Indeed, this perturbation is considered finite; yet it can be large. Hence, the expression for the variation of the action takes the form
\begin{equation}\label{general_action_delta_S}
\delta\mathcal{S} = \int d^{4}x\frac{\sqrt{-g}}{2\kappa}\,\delta g^{\alpha\beta}\left[G_{\alpha\beta}-\kappa T_{\alpha\beta} \right] + \int d^{4}x\frac{\sqrt{-g}}{2\kappa}\nabla_{\mu}\left[g^{\alpha\beta}(\delta\Gamma^{\mu}_{\alpha\beta}) - g^{\alpha\mu}(\delta\Gamma^{\gamma}_{\alpha\gamma})\right] \,. 
\end{equation}
Since $\delta\mathcal{S}$ vanishes for arbitrary variations, we are led to Einstein's equations; however, the second term of $\delta\mathcal{S}$ should not contribute to the field equations, since it contains second derivatives of the metric tensor, therefore, the dynamic equations become at order higher than two. Furthermore, the aforementioned variation is integrated with respect to the natural volume element of the covariant divergence of a vector; then we can apply Stokes's theorem, consequently, we might eliminate this term by evaluating it at the boundary contribution $\partial V$. Indeed, by including the HGY boundary term, this problem is solved \cite{Gibbons:1976ue, York:1972sj}. This extra expression cancels the contributions coming from $g^{\alpha\beta}\delta R_{\alpha\beta}$. Nevertheless, if there were any relevant physical phenomena, it is immediately removed. Instead of cancelling these boundary expressions, an alternative proposal is to take them into account as a physical source of geometric nature \cite{Ridao:2015oba, Ridao:2014kaa}. Under this premise, we will explore how this will bring about a new description of such a scenario. To proceed with this task, let us define the variation of the Ricci tensor as \cite{Ridao:2015oba, Ridao:2014kaa}:
\begin{equation}\label{PHI}
g^{\alpha\beta}\delta R_{\alpha\beta} \equiv \nabla_{\mu}\delta W^{\mu} = \delta\Phi(x^{\mu}) \,,    
\end{equation}
where $\delta W^{\mu}=g^{\alpha\beta}(\delta\Gamma^{\mu}_{\alpha\beta}) - g^{\alpha\mu}(\delta\Gamma^{\gamma}_{\alpha\gamma})$ is a geometric tetra-vector that depends on both the metric tensor and its variation; therefore, $\delta\Phi(x^{\mu})=\delta\Phi$ is a geometric scalar field that emerges as the divergence of this tetra-vector $\delta W^{\mu}$, also it takes into account the back-reaction effects due to the boundary contribution $\partial V$, and it represents a relativistic flow across the border. Furthermore, $\delta\Phi$ becomes zero when the manifold has no boundary $\delta\Phi=0$. Then, we consider the condition:
\begin{equation}\label{condition_1}
\delta\Phi = g_{\alpha\beta}\delta g^{\alpha\beta}\,\Lambda(x) = -g^{\alpha\beta}\delta g_{\alpha\beta}\,\Lambda(x) \,.
\end{equation}
where $\Lambda(x)=\Lambda$ becomes the {\it{Cosmological Boundary Flux Parameter}} (CBFP), and generally depends on the coordinates $x^{\alpha}$. Therefore, to $\delta\mathcal{S}=0$, in eq.~(\ref{general_action_delta_S}), we shall obtain:
\begin{equation}\label{Einstein-eqs}
G_{\alpha\beta} + \Lambda \, g_{\alpha\beta} =  \kappa T_{\alpha\beta} \,.
\end{equation}
We have indeed obtained the Einstein field equation together with the CBFP, which now is no longer taken primordially as a constant. Thus, considering fluctuations (as a geometric response to some physical field fluctuations) $\delta g_{\alpha\beta}$ as the origin of the fluctuations of the curvature: $\delta R_{\alpha\beta} = - \Lambda(x)\,\delta g_{\alpha\beta}$, we obtain the Einstein equations with dynamical $\Lambda(x)$. Consequently, {\it{the flow of the fluctuations of some physical field would be the origin of the cosmological constant on large (cosmological) scales}.} In other words, $\Lambda(x)$ appears as a response to the inclusion of a finite boundary in the Lagrangian formulation of GR. Eqs.~(\ref{condition_1}, \ref{Einstein-eqs}) represent the entire dynamic system, which includes the Einstein equation~(\ref{Einstein-eqs}) and the boundary contribution equation~(\ref{condition_1}). Moreover, this border term can be absorbed into the stress energy tensor, providing an additional energy component in the full description. Thus, the covariant derivative 
\begin{equation}\label{covariant_D}
\nabla_{\beta}\,T^{\alpha\beta} = \frac{1}{\kappa}\,g^{\alpha\beta}\nabla_{\beta}\Lambda \,,
\end{equation}
yields a physical-sourced equation of energy conservation. This means that the flux due to the boundary term becomes the source of the matter sector.  

%
%%%%%%%%%%%%%%%%%%%%%%%%%%%%%%%%%%%%%%%%%%%%%%%%%%%%%
%%%%%%%  BAROTROPIC COSMOLOGICAL PARAMETER  %%%%%%%%%
%%%%%%%%%%%%%%%%%%%%%%%%%%%%%%%%%%%%%%%%%%%%%%%%%%%%%
%
\section{Barotropic cosmological parameter}\label{BCP}
\begin{figure}[htbp] 
\includegraphics[scale=0.75]{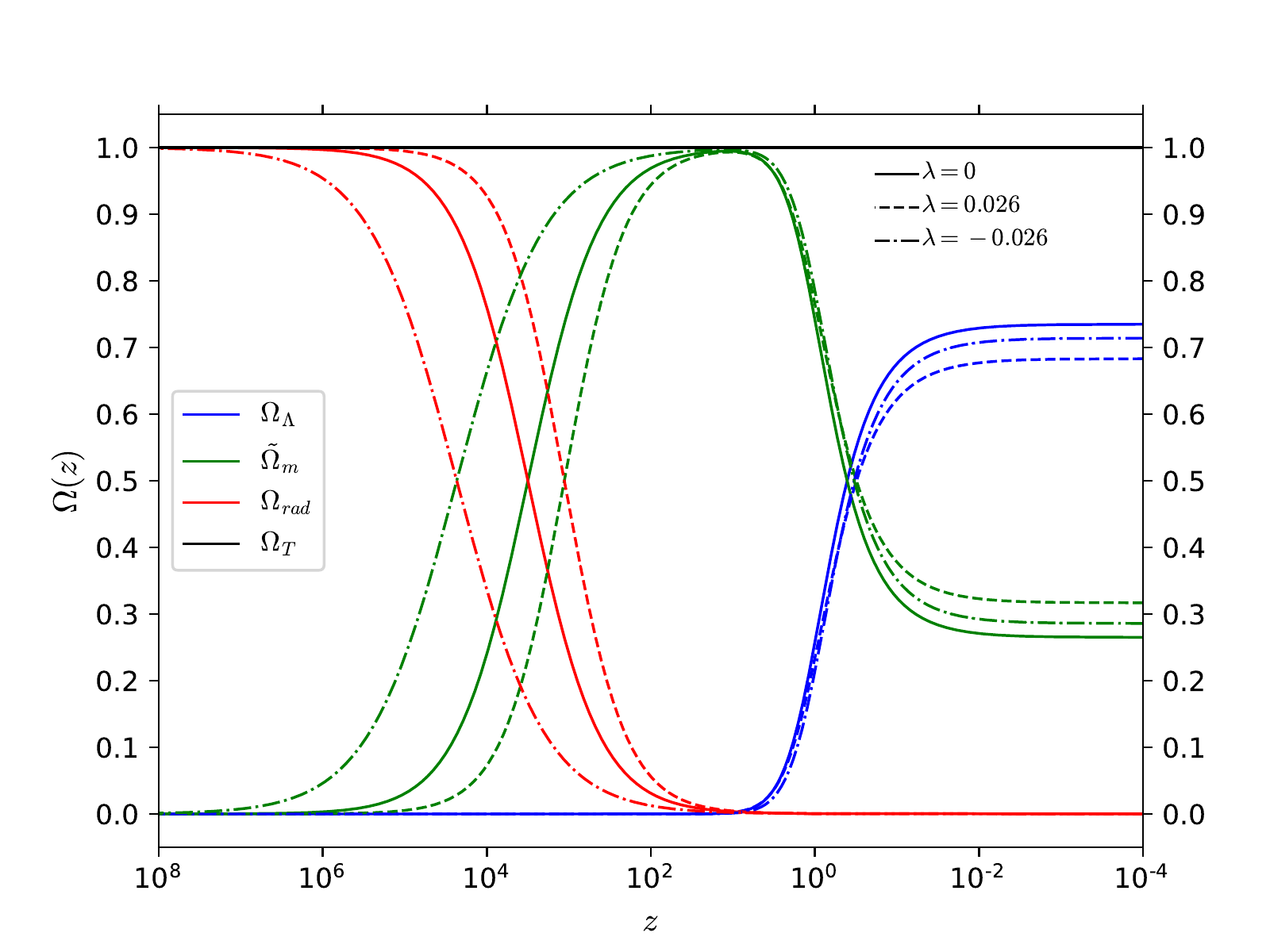}
\caption{Cosmological evolution of $\Omega_{\Lambda}(z)$ (blue), $\tilde{\Omega}_{m}(z)$ (green), $\Omega_{rad}(z)$ (red), and $\Omega_{T}(z)=\sum_{i}\Omega_{i} = 1$ (black). Solid lines correspond to $\lambda=0$ ($\Lambda$CDM); dashed ones for $\lambda=0.026$; and dashed dots when $\lambda=-0.026$.}\label{fig:Omegas}
\end{figure}

We will work within a model described by a perfect fluid with a flat homogeneous and isotropic Friedmann-Lema\^{i}tre-Robertson-Walker (FLRW) metric: 
\begin{equation}\label{RW1}
ds^{2}= -c^{2}dt^{2}+a(t)^{2}\delta_{ij}dx^{i}dx^{j}\,,
\end{equation}
where $t$ is the cosmological time, $a=a(t)$ is the scale factor. For a comoving observer, the components of the relativistic velocity are: $U^{0} = -c \,, U^{1}=U^{2}=U^{3} = 0$. Then, the stress–energy tensor for a perfect fluid takes the form: 
\begin{equation}
T_{\alpha\beta} = \left(\rho + \frac{p}{c^{2}} \right) U_{\alpha}U_{\beta} + p\,g_{\alpha\beta} \,,
\end{equation}
where $\rho$ and $p$ are, respectively, the energy density and pressure for each energy-matter element. The dynamic equations are as follows: 
\begin{eqnarray}
\left(\frac{\dot{a}}{a}\right)^{2} = H^{2} &=& \frac{c^{4}\kappa}{3}\left(\rho_{rad}+\rho_{m}\right) + \frac{c^{2}\Lambda}{3} \,, \label{Hubble_eq}\\
2\frac{\ddot{a}}{a} +  \left(\frac{\dot{a}}{a}\right)^{2} = 2\dot{H} + 3H^{2} &=& -c^{2}\kappa\left(p_{rad}+p_{m}\right) + c^{2}\Lambda \,, \label{dotH_eq} \\
\dot{\rho}_{rad} + 3H\rho_{rad} + \frac{3p_{rad}}{c^{2}}H = 0 \,, && 
\dot{\rho}_{m} + 3H\rho_{m} + \frac{3p_{m}}{c^{2}}H = -\frac{\dot{\Lambda}}{c^{2}\kappa} \,,
\end{eqnarray}
where $H=\dot{a}/a$ is the expansion rate or the Hubble parameter. We consider the matter sector $\rho_{m}=\rho_{b}+\rho_{cdm}$ as the sum of baryons $b$ and cold dark matter $cdm$. Although the radiation density is given by contributions of photons $\gamma$ and ultra–relativistic neutrinos $\nu$, therefore $\rho_{rad}=\rho_{\gamma}+\rho_{\nu}$. For a barotropic component, the equation of state becomes $p=\omega c^{2}\rho$, where $\omega$ is the barotropic parameter. We consider baryons ($b$) and cold dark matter ($cdm$) behave as dust, with vanishing pressure $p_{b} = p_{cdm} = 0$, hence $p_{m}=0$; whilst for radiation we have $p_{rad} = c^{2}\rho_{rad}/3$. Moreover, we take the ansatz
{\footnote{Indeed, in~\cite{Corral:2020lxt,LinaresCedeno:2020uxx} authors proposed a very similar model, they studied Diffusion Models, and in particular the called Barotropic model which is characterised by the diffusion function $Q$; however, our ansatz presents a straightforward interpretation of a given energy transfer between the cosmological boundary flux parameter and the matter sector.}}%
: 
\begin{equation}\label{Lambda}
\Lambda = \Lambda_{0}\left( 1 + \lambda \frac{\rho_{m}}{\rho_{0\,m}} \right) \,,
\end{equation}
where $\rho_{0\, m}$ is the current value of the matter density; and $\lambda$ is a dimensionless constant that controls the coupling of the cosmological parameter with the barotropic energy density, which will be fixed by observational data. Note that our particular ansatz leads to a direct relation between the CBFP and the matter sector. Given the sign of $\lambda$, two distinct physical scenarios manifest: when $\lambda>0$ the former gives its energy to the latter; whilst for $\lambda<0$ the opposite case occurs. Furthermore, the standard $\Lambda$CDM scenario is recovered when $\lambda=0$. Then, the conservation of energy equation becomes:
\begin{equation}
\left(1+\frac{\lambda\Lambda_{0}}{c^{2}\kappa\rho_{0\,m}}\right)\dot{\rho}_{m} + 3H\rho_{m} = 0   \,, 
\end{equation}
from which we get:
\begin{equation}
\frac{\rho_{m}}{\rho_{0\,m}} = \left(\frac{a}{a_{0}}\right)^{-\chi_{m}} \,,\quad \chi_{m} = \frac{3}{\alpha_{m}} \,,\qquad \alpha_{m}=1+\frac{\lambda\Lambda_{0}}{c^{2}\kappa\rho_{0\,m}} = 1+\lambda\left(\frac{\Omega_{0\,\Lambda}}{\Omega_{0\,m}}\right) \,,
\end{equation}
here, $a_{0}$ is the present value of the scale factor, $\Omega_{0\,i}=c^{4}\kappa\rho_{0\,i}/(3H_{0}^{2})$, $\Omega_{0\,\Lambda}=c^{2}\Lambda_{0}/(3H_{0}^{2})$, and $H_{0}$ is the Hubble parameter at present. Since $(a/a_{0})=(1+z)^{-1}$, where $z$ is the redshift, we have for the energy density:
\begin{equation}
\frac{\rho_{m}}{\rho_{0\,m}} = \left(1+z\right)^{\chi_{m}} \,. 
\end{equation}  
The energy density of $\Lambda$ comprises a combination of non-relativistic matter and dark energy components, that is, both elements form a fluid that dominates evolution in distant times, so the correct equation of state is calculated with both constituents: $\rho_{\Lambda}^{(\rm CBFP)} = \rho_{0\,\Lambda}(1 + \lambda\,\rho_{m}/\rho_{0\,m})$, where $\rho_{0\,\Lambda}=\Lambda_{0}/(c^{2}\kappa)$. At the same time, for an accelerated universe $p<0$, and since the matter sector is pressureless, the pressure of $\Lambda$ becomes $p_{\Lambda}^{(\rm CBFP)}=-c^{2}\rho_{0\,\Lambda}$. Therefore, the barotropic parameter is: 
\begin{equation}\label{barotropic_parameter}
\omega_{\Lambda}^{(\rm CBFP)}(z) = \frac{p_{\Lambda}^{(CBFP)}}{c^{2}\rho_{\Lambda}^{(CBFP)}} = -\left(\frac{1}{1+\lambda\left(1+z\right)^{\chi_{m}}}\right) \,.    
\end{equation}
Note that at early times ($z\ll 1$) $\omega_{\Lambda}^{(\rm CBFP)}\rightarrow 0$, that is, the equation of non-relativistic matter; and one also recovers the flat $\Lambda$CDM $\omega_{\Lambda_{0}}=\omega_{DE}=-1$ when $\lambda=0$, where DE stands for Dark Energy. On the other hand, the normalised Friedmann equation can be written as
\begin{equation}\label{Hubblez-dark}
E^{2}(z) \equiv \frac{H^{2}(z)}{H_{0}^{2}} = \Omega_{0\,rad}\left(1+z\right)^{4} + \alpha_{m}\Omega_{0\,m}\left(1+z\right)^{\chi_{m}} + \Omega_{0\,\Lambda} \,.
\end{equation}
Then, in a flat FLRW universe, the Friedman constraint is satisfied during all cosmological evolution, that is, $\sum_{i}\Omega_{i}(z)=1$, where
\begin{equation}
\Omega_{\Lambda}(z) = \frac{\Omega_{0\,\Lambda}}{E^{2}(z)} \,,\quad \Omega_{rad}(z)=\frac{\Omega_{0\,rad}\left(1+z\right)^{4}}{E^{2}(z)} \,,\quad \tilde{\Omega}_{m}(z) = \frac{\alpha_{m}\Omega_{0\,m}\left(1+z\right)^{\chi_{m}}}{E^{2}(z)} \,.
\end{equation}
Moreover, to measure the cosmic acceleration of the expansion of the universe, we use the deceleration parameter, which is given by the relation $q=-\ddot{a}a/\dot{a}^{2}=-1-\dot{H}/H^{2}$, where in terms of the redshift, and $E(z)$ is given by:
\begin{equation}
q(z)=\frac{1}{E^{2}(z)}\left[\Omega_{0\,rad}\left(1+z\right)^{4} - \Omega_{0\,\Lambda} + \frac{1}{2}\left(3-2\alpha_{m}\right) \Omega_{0\,m}\left(1+z\right)^{\chi_{m}}\right] \,. 
\end{equation}
We can exemplify the feedback process between the CBFP and the matter sector. We give an example by taking the following input numerical values due to our statistical analysis performed in the next section (Sect.~\ref{analysis_results}): for $\lambda=0$: $\Omega_{0\,rad}=8.39\times 10^{-5}$, $\Omega_{0\,m}=0.265$, and $\Omega_{0\,\Lambda}=1-\Omega_{0\,rad}-\Omega_{0\,m}=0.735$; and for $\lambda=\pm 0.026$: $\Omega_{0\,rad}=8.57\times 10^{-5}$, $\Omega_{0\,m}=0.299$, and $\Omega_{0\,\Lambda}=(1-\Omega_{0\,rad}-\Omega_{0\,m})/(1+\lambda)=0.683\,(\lambda=+0.026),\, 0.701\,(\lambda=-0.026)$. Note that this value is obtained from the normalised Friedmann equation at $z=0$, that is, $E^{2}(z=0)=1=\Omega_{0\,rad} + \alpha_{m}\Omega_{0\,m} + \Omega_{0\,\Lambda}$, therefore $\Omega_{0\,\Lambda}=(1-\Omega_{0\,rad}-\Omega_{0\,m})/(1+\lambda)$. Fig.~\ref{fig:Omegas} shows this interaction between these two ingredients. We have chosen a particular $\lambda$ with positive and negative signs to illustrate distinct instances: when $\lambda>0$ then $\Lambda$ acts as a source to the matter sector; whilst $\lambda<0$ then $\Lambda$ appears as a friction term, that is, it extracts energy from matter. Note that a positive $\lambda$ yields a smaller dark energy $\Omega_{0\,\Lambda}$ present contribution compared to $\lambda=0$, albeit a larger $\Omega_{0\,m}$; while a negative $\lambda$ closes this gap to the standard $\Lambda$CDM predictions. Hence, to describe a scenario that departs from the $\Lambda$CDM framework at $z=0$, a positive $\lambda$ is preferred, which in turn suggests that the matter sector is nourished by the CBFP. Moreover, the redshift for the matter-radiation equality epoch $z_{eq}$ depends on $\lambda$. For $\lambda=0$ ($\Lambda$CDM) we have $z_{eq}\simeq 3158$; then for $\lambda=0.026$ we have $z_{eq}\simeq 1132$; and for $\lambda=-0.026$ gives $z_{eq}\simeq 24827$. Thus, these outcomes, in fact, produce different implications in cosmic history, from which the positive $\lambda$ example yields the earliest $z_{eq}$, which, in fact, almost coincides with the redshift of the last scattering surface $z_{CMB}\simeq 1100$.        

\begin{figure}[htbp] 
\includegraphics[scale=0.75]{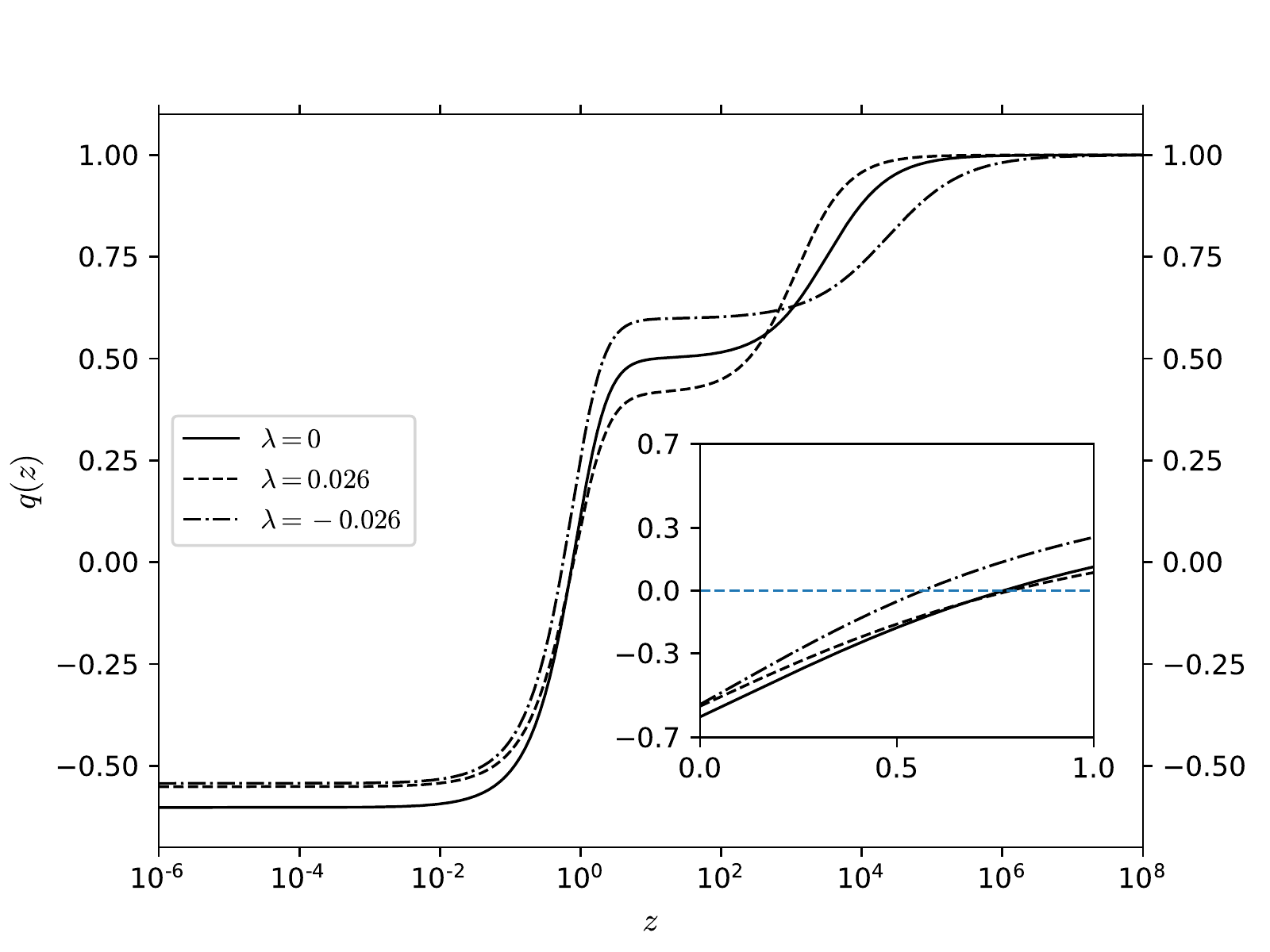}
\caption{Evolution of the deceleration parameter $q(z)$ in terms of the redshift $z$. Solid lines correspond to $\lambda=0$ ($\Lambda$CDM); then dashed ones for $\lambda=0.026$; and dashed dots when $\lambda=-0.026$. The blue dashed line draws the acceleration-deceleration transition epoch at $q=0$.}\label{fig:q}
\end{figure}

Then, fig.~\ref{fig:q} shows the evolution of the deceleration parameter $q(z)$ in terms of the redshift $z$. The current value of $q$ at $z=0$ for various $\lambda$'s is, in fact, almost the same. Having $q_{\Lambda\rm CDM}\simeq -0.60$, then $q_{\lambda>0}\simeq -0.55$, and $q_{\lambda<0}\simeq -0.54$. However, the value of the acceleration-deceleration transition redshift estimate at $q=0$ changes for each scenario. When $z_{\Lambda\rm CDM}\simeq 0.7695$, then $z_{\lambda>0}\simeq 0.7875$, and $z_{\lambda<0}\simeq 0.5745$. Indeed, the negative $\lambda$ result approximates to the transition redshift $z_{t}=0.64^{+0.11}_{-0.07}$ reported in \cite{Moresco:2016mzx}. Similar outcomes are presented in \cite{Planck:2015fie,Riess:2006fw, Lima:2012bx, Busca:2012bu, Capozziello:2015rda, Capozziello:2014zda, Farooq:2013hq, Farooq:2013eea, Rani:2015lia}. 

We purposely selected our numerical input values to illustrate the back-reaction process between the CBFP and the matter sector; which in fact yielded new attractive results. Accordingly, this particular choice is merely the upshot of the statistical analysis performed in order to infer the most likely values of such parameters in light of current late-time astrophysical data. In Sections~\ref{cosmo-constraints} and~\ref{analysis_results} we will detail the entire process. 

\begin{figure}[htbp] 
\includegraphics[scale=0.75]{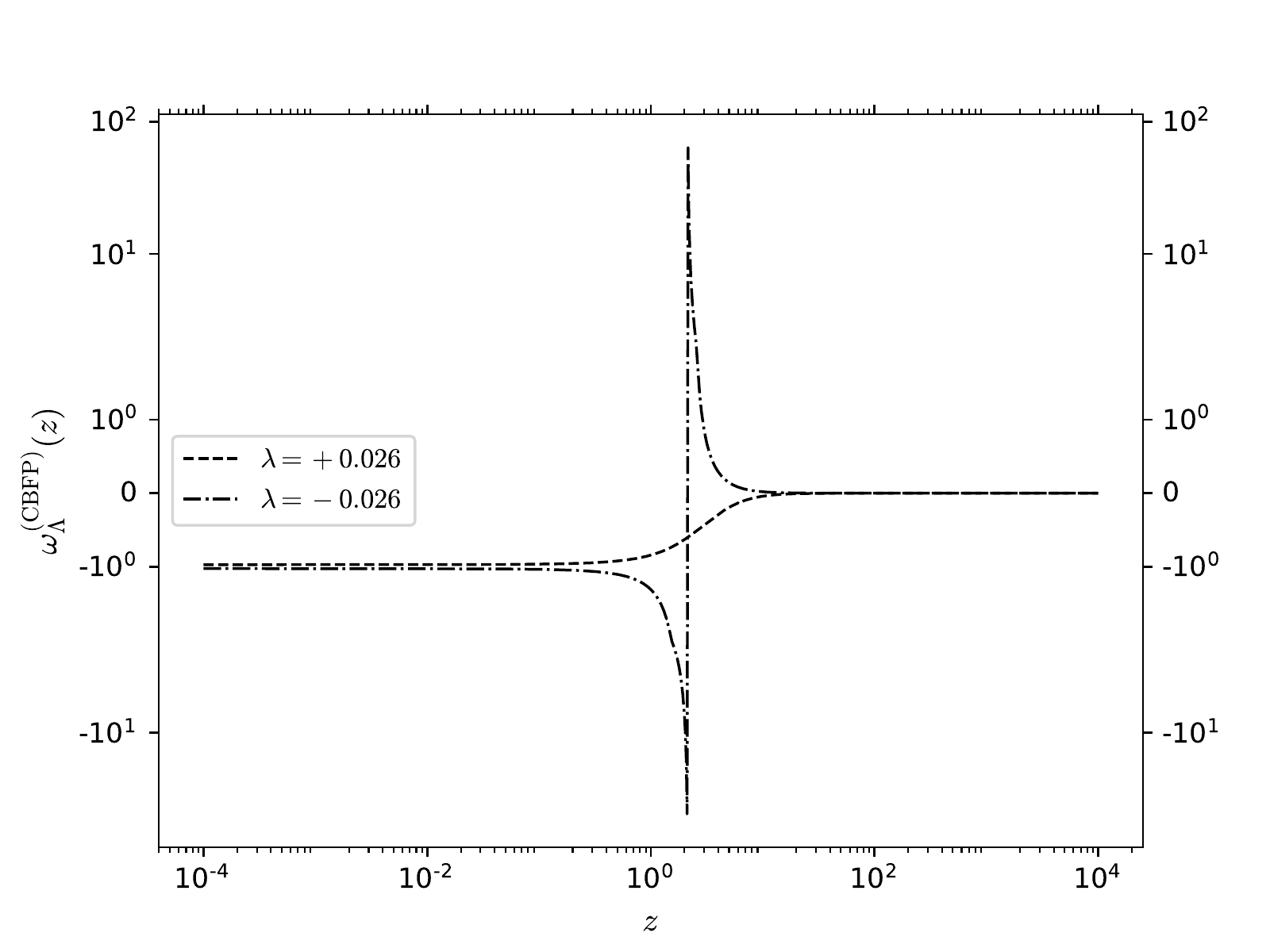}
\caption{Evolution of the barotropic parameter $\omega_{\Lambda}^{(\rm CBFP)}(z)$ in terms of the redshift $z$. The dashed line corresponds to the value $\lambda=0.026$; and the dashed dots one when $\lambda=-0.026$. Note that $\omega_{\Lambda}^{(\rm CBFP)}(\lambda>0)\simeq -0.975$ (Quintessence), and $\omega_{\Lambda}^{(\rm CBFP)}(\lambda<0)\simeq -1.027$ (Phantom). Furthermore, at early times ($z\ll 1$) $\omega_{\Lambda}^{(\rm CBFP)}\rightarrow 0$, that is, the equation of non-relativistic matter component.}\label{fig:w}
\end{figure}

Finally, in this section, we show the evolution of $\omega_{\Lambda}^{(\rm CBFP)}(z)$ with respect to the redshift $z$ (see fig.~\ref{fig:w}). First, the current values of $\omega_{\Lambda}^{(\rm CBFP)}$ at $z=0$ for various $\lambda$'s are: $\omega_{\Lambda}^{(\rm CBFP)}(\lambda>0)\simeq -0.975$, and $\omega_{\Lambda}^{(\rm CBFP)}(\lambda<0)\simeq -1.027$. The positive value lies in the Quintessence regime ($\omega_{\rm DE}>-1$)~\cite{Tsujikawa:2013fta}, whilst the negative one sits in the Phantom zone ($\omega_{\rm DE}<-1$)~\cite{Ludwick:2017tox}. Moreover, $\omega_{\Lambda}^{(\rm CBFP)}\simeq \omega_{m} = 0$ at $z\ll 1$, hence one recovers the barotropic parameter of nonrelativistic matter at early times.   

%

%
%%%%%%%%%%%%%%%%%%%%%%%%%%%%%%%%%%%%%%%%%%%%%%%%%%%%%
%%%%%%%  COSMOLOGICAL CONSTRAINTS  %%%%%%%%%%%%%%%%%%
%%%%%%%%%%%%%%%%%%%%%%%%%%%%%%%%%%%%%%%%%%%%%%%%%%%%%
%
\section{Cosmological constraints}\label{cosmo-constraints}
In this section, we constrain the free parameters of the model $\Theta =\{h, \Omega_{0\, m}, \lambda \}$, where $h=H_{0}/100$. To achieve this task, a merit function $\log \mathcal{L} \sim$ $\chi^2$ is minimised by using late-time Observational Hubble Data (OHD) and Type Ia Supernovae (SNe Ia) distance modulus. Then, we compute the best fit value by means of the affine-invariant Markov Chain Monte Carlo method (MCMC) \cite{2010CAMCS...5...65G}. To compute posterior probabilities, we use Cobaya software \cite{Torrado:2020dgo}, which is a general-purpose Bayesian analysis code. Note that we estimate the free parameters and their confidence regions via a Bayesian statistical analysis; then we apply the Gaussian likelihood function:
\begin{equation}\label{distributions}
\mathcal{L}_I \sim \exp\left(-\frac{\chi^2_I}{2} \right)\,,
\end{equation}
here, $I$ stands for each data set under consideration, namely OHD, SNe Ia; and their joint analysis with $\chi^{2}_{\rm joint}=\chi^{2}_{\rm OHD}+\chi^{2}_{\rm SNe}$. They are described in the subsequent segments, along with their corresponding $\chi^{2}$ functions. We compute $\Omega_{0\,rad}=2.469\times 10^{-5}h^{-2}(1+0.2271\,N_{eff})$ \cite{2011ApJS..192...18K, Magana:2017nfs}, where $N_{eff}=3.04$ is the standard number of relativistic species \cite{Mangano:2001iu}. Also, from the normalised Friedmann equation at $z=0$, that is, $E^{2}(z=0)=1=\Omega_{0\,rad} + \alpha_{m}\Omega_{0\,m} + \Omega_{0\,\Lambda}$, we obtain $\Omega_{0\,\Lambda}=(1-\Omega_{0\,rad}-\Omega_{0\,m})/(1+\lambda)$. Moreover, since we have no previous knowledge about the parameters to analyse, we have considered flat priors, since they are the most conventional to use. They are $\Omega_{0\, m}\in [0.1,0.5]$; $h\in [0.6,0.76]$; and $\lambda\in [-1,1]$. Furthermore, to monitor the convergence of the posteriors, we employ the Gelman–Rubin criterion~\cite{10.2307/2246093}, $R-1$. We have selected $R-1<1.0\times 10^{-2}$ for both the $\Lambda$CDM and CBFP scenarios. 
%
%%%%%%%%%%%%%%%%%%%%%%%%%%%%%%%%%%%%%%%%%%%%%%%%%%%%%
%%%%%%%  OBSERVATIONAL HUBBLE DATA  %%%%%%%%%%%%%%%%%
%%%%%%%%%%%%%%%%%%%%%%%%%%%%%%%%%%%%%%%%%%%%%%%%%%%%%
%
\subsection{Observational Hubble Data}
We calculate the optimal model parameter, $H_0$, by minimising the merit function:
\begin{equation}\label{chiOHD}
\chi_{\rm OHD}^2=\sum_{i=1}^{N_H} \left(\frac{H_{th}(z_i, \Theta(h, \Omega_{0\, m}, \lambda  ))-H_{obs}(z_i)}{\sigma_{obs}(z_{i})} \right)^2 \,,
\end{equation}
where $H_{th}$ and $H_{obs}$ are the theoretical and observational Hubble parameters at redshift $z_{i}$, respectively; then $\sigma_{obs}(z_{i})$ is the associated error of $H_{obs}(z_{i})$; and $\Theta(h,\Omega_{0\, m},\lambda)$ denotes the free parameter space of $H_{th}$ (eq.~\eqref{Hubblez-dark}). The sample consists of $N_{H}=52 \, H(z)$ measurements in the redshift range $0.0 < z< 2.36$ \cite{Magana:2017nfs}. These data comes from Baryon Acoustic Oscillations (BAO)~\cite{2011MNRAS.416.3017B, 10.1093/mnras/stx721, 10.1093/mnras/sty506, deSainteAgathe:2019voe, Blomqvist:2019rah} and Cosmic Chronometers~\cite{Jimenez:2001gg}. 
%
%%%%%%%%%%%%%%%%%%%%%%%%%%%%%%%%%%%%%%%%%%%%%%%%%%%%%
%%%%%%%%%%%%%%%%%  SNIa SUPERNOVAE  %%%%%%%%%%%%%%%%%
%%%%%%%%%%%%%%%%%%%%%%%%%%%%%%%%%%%%%%%%%%%%%%%%%%%%%
%
\subsection{Supernovae Ia: SNe Ia}
Data from SNe Ia observations is usually released as a distance modulus $\mu$. In our study, we will use the compilation of observational data for $\mu$ given by the Pantheon Type Ia catalogue \cite{Pan-STARRS1:2017jku}, which consists of $N_{\mu}=1048$ SNe data samples, which includes observations up to redshift $z=2.26$. The model for the observed distance modulus $\mu$ is~\cite{Pan-STARRS1:2017jku, Kessler:2016uwi}: 
\begin{equation}\label{mu_proposed}
\mu = m_{b} - \mathcal{M} \,,
\end{equation}
where $m_b$ is the apparent B-band magnitude of a fiducial SNe Ia, and $\mathcal{M}$ is a nuisance parameter, which in fact is strongly degenerated with respect to $H_{0}$ \cite{Pan-STARRS1:2017jku}. To overcome this problem, we will follow the BEAMS method proposed in \cite{Pan-STARRS1:2017jku, Kessler:2016uwi}. First, the theoretical distance modulus in a flat FLRW geometry is given by: 
\begin{equation}
\mu_{th}(z_i, \Theta)=5\log_{10}\left(\frac{d_L(z_i, \Theta)}{Mpc}\right)+\bar{\mu} \,, \quad \bar{\mu}=5\left[\log_{10}{\left(c\right)}+5\right]
\end{equation}
where $c$ is the speed of light given in units of $\rm km\,s^{-1}$, and $d_L(z_i, \Theta)$ is the luminosity distance: 
\begin{equation}
d_L(z_i, \Theta)=\frac{(1+z_{i})}{H_0}\int_0^{z_{i}}\frac{dz'}{E(z',\Theta)} \,.
\end{equation}
This relation allows us to contrast our theoretical model with respect to the observations by minimising the merit function:
\begin{equation}\label{chiSNe}
\chi_{\rm SNe}^2=\sum_{i=1}^{N_{\mu}} \left(\frac{\mu_{th}(z_i, \Theta)-\mu_{i}}{\sigma_{i}} \right)^2 \,,
\end{equation}
where $\mu_i$ and $\mu_{\rm th}$ are the observational and theoretical distance modulus of each SNe Ia at redshift $z_i$, respectively; $\sigma_i$ is the error in the measurement of $\mu_i$, and $\Theta=\{h, \Omega_{0\, m}, \lambda \}$ represents all the free parameters of the respective model. However, we will follow another method to simplify our analysis. First, eq.~\eqref{chiSNe} can be written in matrix notation (bold symbols): 
\begin{equation}\label{chiSNematrix}
\chi^{2}_{\rm SNe} = \mathbf{M}^\dagger\mathbf{C}^{-1}\mathbf{M},
\end{equation}
where $\mathbf{C}$ is the total covariance matrix given by:
\begin{equation}\label{covariancematrix}
\textbf{C}=\textbf{D}_{\rm stat}+\textbf{C}_{\rm sys} \,, 
\end{equation}
and $\mathbf{M}=\mathbf{m}_{b}-\mbox{\boldmath$\mu$}_{\rm th}\left(z_{i},\theta\right)-\mbox{\boldmath$\mathcal{M}$}$ \cite{Corral:2020lxt}. The diagonal matrix $\textbf{D}_{\rm stat}$ only contains the statistical uncertainties of $m_{b}$ for each redshift, whilst $\textbf{C}_{\rm sys}$ denotes the systematic uncertainties in the BEAMS with the bias correction approach. We can even simplify this method by reducing the number of free parameters and marginalising over $\mathcal{M}$, we use $\mathcal{M}=\bar{\mathcal{M}}-\bar{\mu}$ with $\bar{\mathcal{M}}$ being an auxiliary nuisance parameter~\cite{Corral:2020lxt}. Moreover, eq.~(\ref{chiSNematrix}) can be expanded as~\cite{Lazkoz:2005sp,Corral:2020lxt}:
\begin{equation}\label{chi2projected}
\chi^{2}_{\rm SNe}=A\left(z,\theta\right)-2B\left(z,\theta\right)\bar{\mathcal{M}}+C\bar{\mathcal{M}}^{2} \,, 
\end{equation}
where
\begin{equation}
A\left(z,\theta\right) = \bar{\mathbf{M}}^{\dagger}\textbf{C}^{-1}\bar{\mathbf{M}} \,,\quad B\left(z,\theta\right) = \bar{\mathbf{M}}^{\dagger}\textbf{C}^{-1}\,\textbf{1} \,,\quad C = \textbf{1}^{\dagger}\, \textbf{C}^{-1}\, \textbf{1} \,, 
\end{equation}
with $\bar{\mathbf{M}}=\mathbf{m}_{B}-\mbox{\boldmath$\mu$}_{\rm th}\left(z_{i},\theta\right)+\bar{\mbox{\boldmath$\mu$}}$~\cite{Corral:2020lxt}. Note that, in fact, $\bar{\mathbf{M}}$ no longer contains any troublesome parameters. Finally, minimising eq.~\eqref{chi2projected} with respect to $\bar{\mathcal{M}}$ gives $\bar{\mathcal{M}}=B/C$ and reduces to:
\begin{equation}\label{chi2SNeprojected}
\chi^{2}_{\rm SNe}\Big|_{\rm min}=A\left(z,\theta\right)-\frac{B\left(z,\theta\right)^{2}}{C}. 
\end{equation}
Both eqs.~(\ref{chi2SNeprojected},\ref{chiSNematrix}) yield the same information; however, eq.~(\eqref{chi2SNeprojected}) only contains the free parameters of the model, and the nuisance term has been marginalised. Thus, we will implement eq.~(\ref{chi2SNeprojected}) as our merit function. The Pantheon data set is available online in the GitHub repository \href{https://github.com/dscolnic/Pantheon}{https://github.com/dscolnic/Pantheon}: the document \textit{lcparam\_full\_long.txt} contains the corrected apparent magnitude $m_{b}$ for each SNe Ia together with their respective redshifts ($z_{i}$) and errors ($\sigma_{i}$); and the file \textit{sys\_full\_long.txt} includes the full systematic uncertainties matrix $\textbf{C}_{\rm sys}$.
%
%%%%%%%%%%%%%%%%%%%%%%%%%%%%%%%%%%%%%%%%%%%%%%%%%%%%%
%%%%%%%%%%% ANALYSIS AND RESULTS  %%%%%%%%%%%%%%%%%%%
%%%%%%%%%%%%%%%%%%%%%%%%%%%%%%%%%%%%%%%%%%%%%%%%%%%%%
%
\section{Analysis and Results}\label{analysis_results}
Data from SNe Ia alone yields bias results due to the nuisance parameter $\mathcal{M}$, therefore, $H_{0}$ cannot be determined using only this set of information. Therefore, to constrain $H_{0}$, we must combine it with other observations. Both $\Lambda$CDM and CBFP models are contrasted with a joint analysis using OHD and SNe Ia data through their corresponding Hubble parameters. Once the model is constrained, we will compare the proposed CBFP model with the $\Lambda$CDM one using the Bayesian Information Criterion (BIC) \cite{BIC} defined as:
\begin{equation}
BIC = \chi_{min}^2+k \ln{N} \,,
\end{equation}
where $\chi_{min}^2$ is log-likelihood of the model, $k$ is the number of free parameters of the optimised model; and $N$ is the number of data samples. This criterion gives us a quantitative value to select among several models. Following Jeffrey-Raftery's \cite{10.2307/271063} guidelines, if the difference in BICs between the two models is 0–2, this constitutes `weak' evidence in favour of the model with the smaller BIC; a difference in BICs between 2 and 6 constitutes `positive' evidence; a difference in BICs between 6 and 10 constitutes `strong' evidence; and a difference in BICs greater than 10 constitutes `very strong' evidence in favour of the model with smaller BIC.

% On the other hand, one can measure the tension among the current constraints on $H_{0}$, namely late and early cosmological and astrophysical observations. We follow the use of the estimator $T_{H0}$ to quantify the discordance or tension in current determinations of~$H_0$ between the Plank+BAO data and our fitting results \cite{Camarena:2018nbr,LinaresCedeno:2020uxx}:
% %
% \begin{equation}\label{TH-stimator}
% T_{H_{0}}=\frac{|H_{0} - H_{0}^{\rm CMB}|}{\sqrt{\sigma_{H_{0}}^{2} + \sigma_{H_{0}\,}^{2\, \rm CMB}}} \ ,,
% \end{equation}
% %
% where $H_0$ and $\sigma^2_{H0}$ are the mean and variance of our fitting parameters; and $H_{0}^{\rm CMB}=67.66 ^{+0.42}_{-0.42}\rm\,\, Km \,s^{-1}\,Mpc^{-1}$ (Planck + BAO) \cite{Planck:2018nkj}. Furthermore, in the case of an asymmetric standard deviation, we implement $\sigma_{H_{0}}=(\sigma_{H_{0}}^{+}+\sigma_{H_{0}}^{-})/2$. According to~\cite{Lin:2017ikq,Trotta:2008qt} the qualitative interpretation of the tension estimator $T_{H_{0}}$ is: $<1.4$ no significant tension; between 1.4 and 2.2 weak tension; then 2.2-3.1 moderate tension; and when $>3.1$ we have strong tension.

\begin{figure}[htbp] 
\includegraphics[scale=0.5]{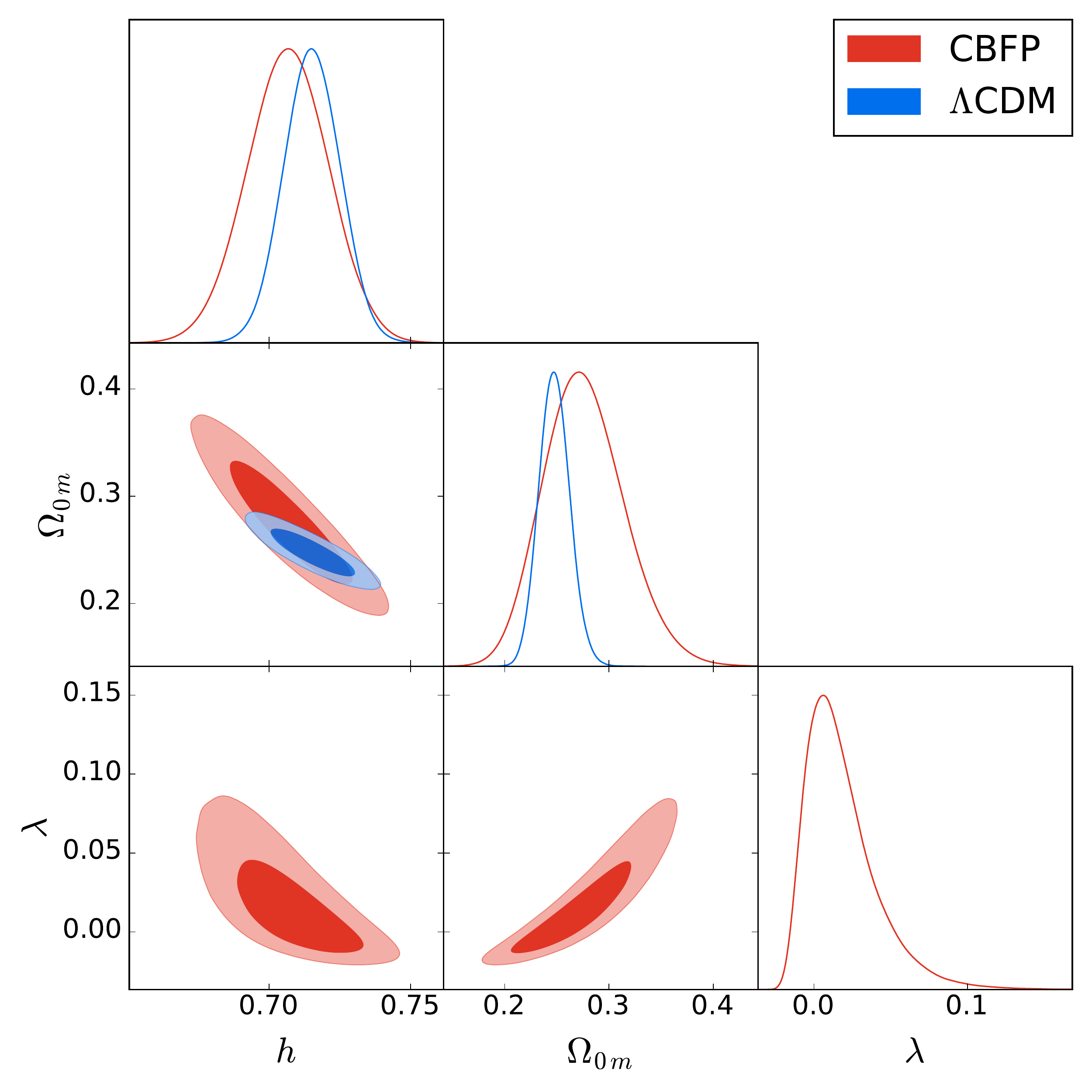}
\caption{OHD data constraints of $h$, $\Omega_{0\,m}$, and $\lambda$ for the flat $\Lambda$CDM model and the CBFP scenario, using Bayesian statistical analysis of Sect.~\ref{cosmo-constraints}. The admissible regions correspond to $1\sigma\left(68.3\%\right)$, and $2\sigma\left(95.5\%\right)$, CL, respectively. In this case, the best-fit values of the CBFP scenario at $1\sigma$ are $h=0.707^{+0.014}_{-0.014}$, $\Omega_{0\,m}=0.276^{+0.034}_{-0.042}$, and $\lambda=0.018^{+0.011}_{-0.028}$.}\label{OHD}
\end{figure}

\begin{figure}[htbp] 
\includegraphics[scale=0.5]{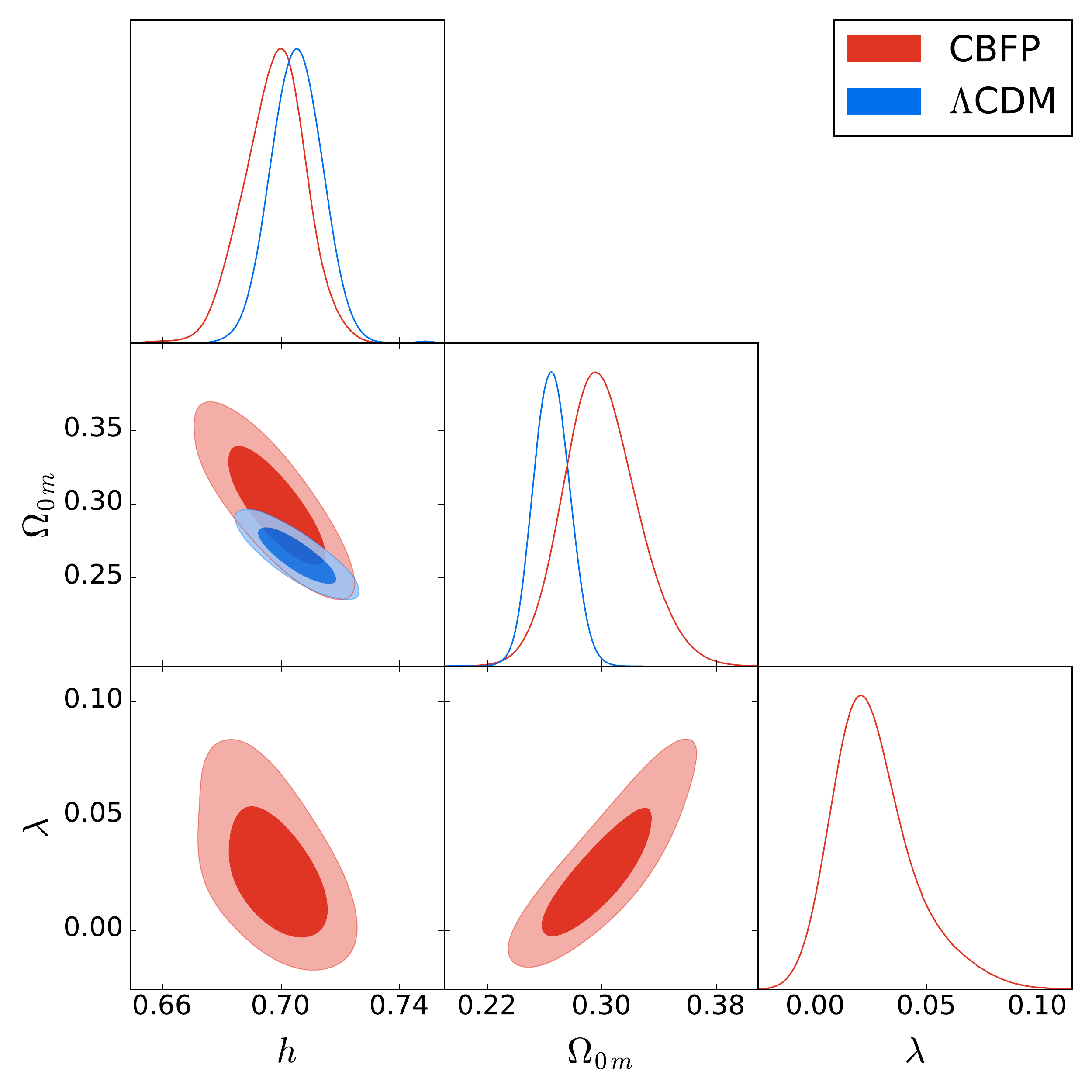}
\caption{Joint constraints of $h$, $\Omega_{0\,m}$, and $\lambda$ for the flat $\Lambda$CDM model and the CBFP scenario, using Bayesian statistical analysis of Sect.~\ref{cosmo-constraints}. The admissible regions correspond to $1\sigma\left(68.3\%\right)$, and $2\sigma\left(95.5\%\right)$, CL, respectively. In this case, the best-fit values of the CBFP scenario at $1\sigma$ are $h=0.698^{+0.011}_{-0.0095}$, $\Omega_{0\,m}=0.299^{+0.024}_{-0.027}$, and $\lambda=0.026^{+0.013}_{-0.021}$.}\label{joint}
\end{figure}

Table~\ref{tab:cobaya} shows the best-fit values for the parameters of each model from the OHD and the joint (OHD + SNe Ia) analysis; together with their $\chi^{2}_{min}$, and BIC indicators. Furthermore, figures~\ref{OHD} (OHD) and~\ref{joint} (joint) show the posteriors of the parameters within the scenarios $\Lambda$CDM (blue) and CBFP (red). The best-fit values for the CBFP model, derived from the joint analysis, are $h=0.698^{+0.011}_{-0.0095}$, $\Omega_{0\,m}=0.299^{+0.024}_{-0.027}$, and $\lambda=0.026^{+0.013}_{-0.021}$; which are, in fact, quite similar to Corral, et. al.~\cite{Corral:2020lxt}; however, differences between the upshots could be due to the fact that we have included radiation in the statistical analysis, and they have not. Moreover, ref~\cite{LinaresCedeno:2020uxx} analysed Diffusion Model as well; nonetheless, it also constrained the parameters with CMB and more late-time data sets; hence, its survey is more complex, but we obtained similar fitted values. It is important to note that both models, $\Lambda$CDM and CBFP, lead to quite similar values of $h$ either using only OHD or OHD + SNe Ia. This upshot indicates that the OHD data~\cite{Magana:2017nfs}, a combination of Cosmic Chronometers~\cite{Jimenez:2001gg} and BAO observations~\cite{2011MNRAS.416.3017B, 10.1093/mnras/stx721, 10.1093/mnras/sty506, deSainteAgathe:2019voe, Blomqvist:2019rah}, have a greater influence on the value of $H_{0}$ ($H_{0}=100\,h\rm\,\, Km \,s^{-1}\,Mpc^{-1}$). In fact, the authors in~\cite{Corral:2020lxt}, using only SNe Ia data, reported the fitted value: $h=0.732^{+0.017}_{-0.017}$ for $\Lambda$CDM and their Diffusion Model. Hence, indicating that in fact the OHD data dominate over the SNe Ia observations; however, the joint analysis yields smaller fitted values than the OHD data themselves. 

Furthermore, the joint CBFP outcome indicates a lower $\chi^{2}_{min}$ than $\Lambda$CDM; nevertheless, given that the latter scenario has fewer free parameters, it gives a lower BIC. Indeed, we obtain $\Delta BIC_{Joint}=4.94$. Although this value implies `positive' evidence in favour of $\Lambda$CDM, there is still room to constrain the parameter space further with additional early and late times data. 

\begin{table}[h!]
\centering
\begin{tabular}{| c | c | c |}
\hline
Data  & \qquad Best-fit values: mean$^{+1\sigma}_{-1\sigma}$ \qquad  & Goodness of fit \\
\hline
 &  $h \qquad\qquad \Omega_{0\, m} \qquad\qquad \lambda$  & $\chi^{2}_{min}$ \quad BIC \\
\hline
\multicolumn{3}{ |c| }{$\Lambda$CDM} \\ 
\hline
OHD  & $0.7151^{+0.0099}_{-0.0099} \quad 0.248^{+0.015}_{-0.015} \quad -- $  & $28.88 \quad 36.79 $\\
\hline
Joint  & $0.7053^{+0.0088}_{-0.0088} \quad 0.265^{+0.013}_{-0.013} \quad -- $  & $1059.59 \quad 1073.60 $\\
\hline
\multicolumn{3}{ |c| }{CBFP} \\ 
\hline
OHD  & $0.707^{+0.014}_{-0.014} \quad 0.276^{+0.034}_{-0.042} \quad 0.018^{+0.011}_{-0.028} $ & $29.22 \quad 41.07  $ \\
\hline
Joint  & $0.698^{+0.011}_{-0.0095} \quad 0.299^{+0.024}_{-0.027} \quad 0.026^{+0.013}_{-0.021} $ & $1057.54 \quad 1078.54  $ \\
\hline
\end{tabular}
\caption{Results of the best-fit parameters and statistical indicators. The uncertainties correspond to $1\sigma$ $(68.3\%)$ of the confidence level (CL). We consider the criterion $R-1<1.0\times 10^{-2}$ for the OHD data and the joint analysis.}
\label{tab:cobaya}
\end{table}

To end this section, we want to stress that the joint analysis of CBFP produces a smaller $H_{0}^{\rm CBFP}=69.80\rm\,\, Km \,s^{-1}\,Mpc^{-1}$ ($H_{0}=100\,h\rm\,\, Km \,s^{-1}\,Mpc^{-1}$) in contrast to the flat $\Lambda$CDM result $H_{0}^{\Lambda\rm CDM}=70.53\rm\,\, Km \,s^{-1}\,Mpc^{-1}$. Indeed, this outcome closes the breach with respect to the CMB value $H_{0}^{\rm CMB}=67.70\rm\,\, Km \,s^{-1}\,Mpc^{-1}$ (Planck + BAO) \cite{Planck:2018nkj}. Despite that, more statistical analysis is needed to make a complete comparison between flat $\Lambda$CDM and CBFP using CMB data. 

% Additionally, the CBFP estimator $T_{H0}=1.93$ eases the Hubble tension, where according to~\cite{Lin:2017ikq,Trotta:2008qt} this only constitutes weak tension, contrary to $T_{H0}=2.94$ of $\Lambda$CDM yielding moderate tension. Thus, this might suggest that the CBFP model is effectively a good candidate to alleviate the $H_{0}$ tension. However, to in fact reduce this conundrum, more statistical analysis needs to be done. 

%
%

%
%%%%%%%%%%%%%%%%%%%%%%%%%%%%%%%%%%%%%%%%%%%%%%%%%%%%%
%%%%%%%%%%%%%%%%%% CONCLUSIONS  %%%%%%%%%%%%%%%%%%%%
%%%%%%%%%%%%%%%%%%%%%%%%%%%%%%%%%%%%%%%%%%%%%%%%%%%%%
%
\section{Conclusions}\label{conclusions}
Our approach pursues the objective of presenting a novel scheme of the origin of the accelerated expansion of the universe. Three key points that we wish the reader would take home. First, we derive the gravitational field equations via the variation of EH action, and when the underlying spacetime manifold has a boundary $\partial V$, the variation to the Ricci tensor evaluated at the border is taken into account as a physical source of geometric nature, namely $g^{\alpha\beta}\delta R_{\alpha\beta} = g_{\alpha\beta}\delta g^{\alpha\beta}\,\Lambda$. Hence, this back-reaction contribution gives rise to the CBFP. Second, we implemented this new approach to a flat FLRW universe, along with a barotropic fluid. We proposed an ansatz for which the main motivation was simply to link $\Lambda$ with the matter sector; therefore, only one additional parameter was introduced: $\lambda$. Third, by statistical analysis, we calculated the best-fit value of $\Theta =\{h, \Omega_{0\, m}, \lambda \}$ using the affine-invariant MCMC. We used late-time data for OHD and SNe Ia. The result of the joint analysis is presented in Table~\ref{tab:cobaya}. For the joint data, we have: $h=0.698^{+0.011}_{-0.095} \,, \Omega_{0\,m} = 0.299^{+0.024}_{-0.027} \,, \lambda = 0.026^{+0.013}_{-0.021}$.  

Moreover, cosmic history might have different backgrounds given by the deceleration parameter and the time of the matter-radiation equality epoch $z_{eq}$. Remarkably for $\lambda=0.026$ we have $z_{eq}\simeq 1132$, which, in fact, almost coincides with the redshift of the last scattering surface $z_{CMB}\simeq 1100$. Together with the value of the acceleration-deceleration transition redshift estimate at $q=0$, where $\lambda=-0.026$ yields $z_{t}\simeq 0.5745$, and this value approximates the transition redshift $z_{t}=0.64^{+0.11}_{-0.07}$ \cite{Moresco:2016mzx}. Furthermore, the barotropic parameter lies in the Quintessence regime $\omega_{\Lambda}^{(\rm CBFP)}(\lambda=0.026)\simeq -0.975$. 

When comparing the statistical results of CBFP with those of $\Lambda$CDM, $\chi^{2}_{min}$ is lower; however, a model with more free parameters is penalised with a larger BIC, this being the case for the CBFP scenario, which, for instance, has $\Delta BIC_{Joint}=4.94$. Although this outcome suggests that $\Lambda$CDM might be `positive' compared to CBFP, there is still room to constrain the parameter space further with additional early- and late-time data sets. Nonetheless, we want to emphasise that the joint analysis produces a smaller $H_{0}=69.8\rm\,\, Km \,s^{-1}\,Mpc^{-1}$ in contrast to the $\Lambda$CDM result $H_{0}=70.53\rm\,\, Km \,s^{-1}\,Mpc^{-1}$. Indeed, this outcome approaches the CMB value more $H_{0}^{\rm CMB}=67.70\rm\,\, Km \,s^{-1}\,Mpc^{-1}$ (Planck + BAO) \cite{Planck:2018nkj}. However, to make a full comparison between flat $\Lambda$CDM and CBFP we must use CMB data as well. Moreover, it is well known that $H_{0}$ and the dark energy equation of state $\omega_{DE}$ are anti-correlated~\cite{Vagnozzi:2019ezj, Alestas:2020mvb, Lee:2022cyh}; that is, for a flat $\Lambda$CDM we have $\omega_{DE}=-1$, so if $H_{0}^{\rm CBFP}<H_{0}^{\Lambda\rm CDM}$ then $\omega_{DE}>-1$, and, in fact, we have $\omega_{\Lambda}^{(\rm CBFP)}(\lambda=0.026)\simeq -0.975>-1$. Thus, the anti-correlation between $H_{0}$ and $w_{DE}$ provides us with an explanation of the lower value of $H_{0}^{\rm CBFP}$ in the CBFP model versus the flat $\Lambda$CDM $H_{0}^{\Lambda\rm CDM}$. 

Certainly, this job opens up new paths to explore. We can name at least three routes to search for. First, we assumed that the CBFP couples only to the matter sector, but one can extend this premise to other sectors as well. Second, there are still more important observational data sets to be considered in future research; for instance, the current Planck + BAO likelihoods; and more late-time surveys. Third, we must study the observable consequences of cosmological perturbations on the large structure formation or the CMB.

%
%%%%%%%%%%%%%%%%%%%%%%%%%%%%%%%%%%%%
%%%%%%%%%%%   APPENDIX   %%%%%%%%%%%
%%%%%%%%%%%%%%%%%%%%%%%%%%%%%%%%%%%%
%
\appendix
\section{Negative contribution to the matter sector}\label{appendix_a}
\begin{figure}[htbp] 
\includegraphics[scale=0.5]{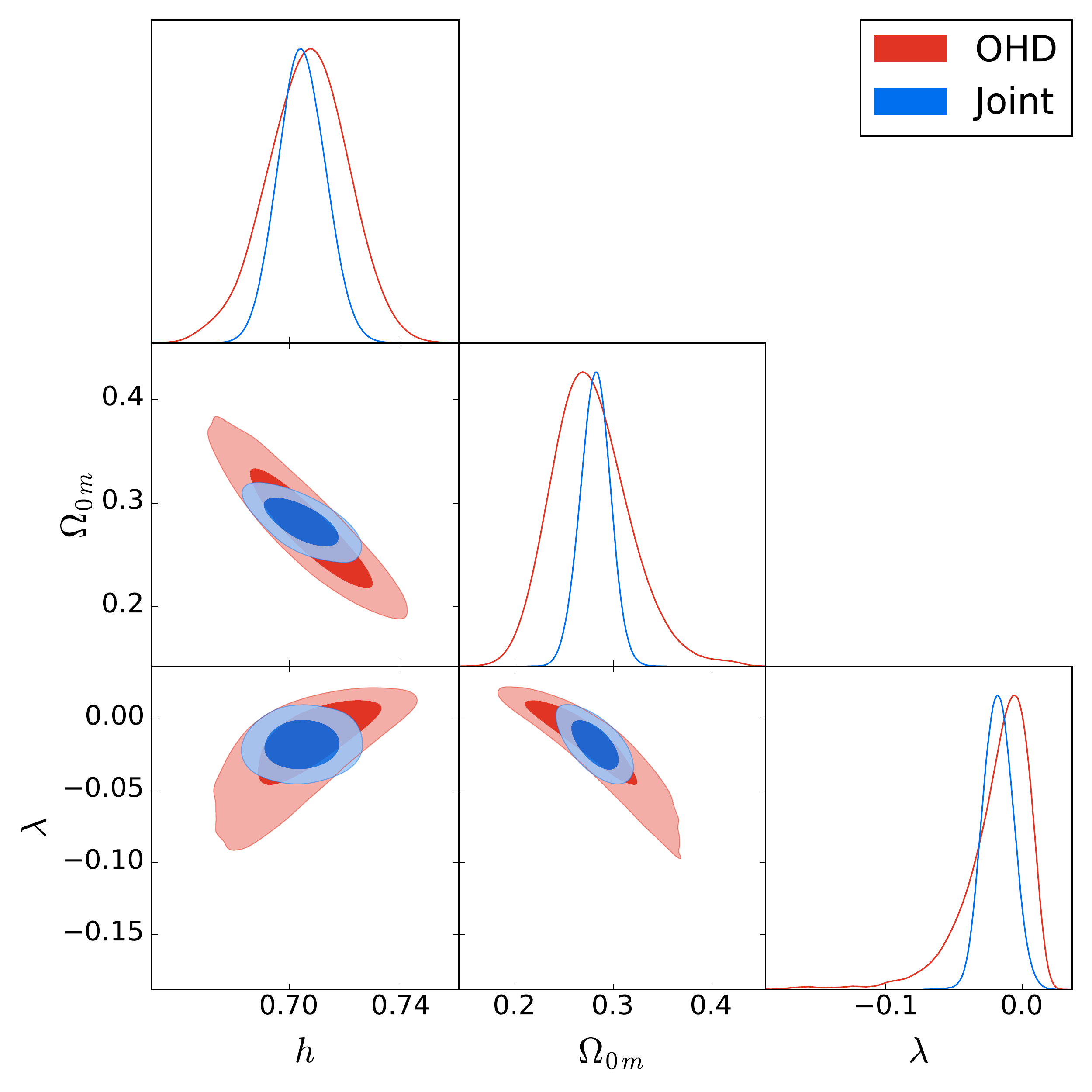}
\caption{Joint (blue) and marginalised OHD (red) constraints of $h$, $\Omega_{0\,m}$, and $\lambda$ for the CBFP scenario with a negative contribution to the matter sector, using Bayesian statistical analysis of Sect.~\ref{cosmo-constraints}. The admissible regions correspond to $1\sigma\left(68.3\%\right)$, and $2\sigma\left(95.5\%\right)$, CL, respectively. In this case, the best-fit values obtained from the joint analysis at $1\sigma$, are $h=0.704^{+0.0085}_{-0.0085}$, $\Omega_{0\,m}=0.282^{+0.015}_{-0.015}$, and $\lambda=-0.018^{+0.011}_{-0.011}$.}\label{joint_negative_lambda}
\end{figure}

In this appendix we include the same Bayesian statistical analysis done in Sect.~\ref{cosmo-constraints}; however, we change the sign of the second term of eq.~(\ref{Lambda}), so now we have:
\begin{equation}\label{NLambda}
\Lambda = \Lambda_{0}\left( 1 - \lambda \frac{\rho_{m}}{\rho_{0\,m}} \right) \,.
\end{equation}
This time Table~\ref{tab:cobaya_negative} shows the best-fit values for the parameters of the CBFP model with a negative contribution to the matter sector from the OHD and the joint (OHD + SNe Ia) analysis; together with their $\chi^{2}_{min}$, and BIC indicators. Also, fig.~\ref{joint_negative_lambda} shows the joint (blue) and marginalised OHD (red) constraints of $h$, $\Omega_{0\,m}$, and $\lambda$ for the CBFP scenario. In this case, the best-fit values obtained from the joint analysis at $1\sigma$, are $h=0.704^{+0.0085}_{-0.0085}$, $\Omega_{0\,m}=0.282^{+0.015}_{-0.015}$, and $\lambda=-0.018^{+0.011}_{-0.011}$. Note that both the OHD data and the joint analysis yield almost the same fitted values, and, in fact, since $\lambda<0$ we basically recover the previous results with $\lambda>0$. However, this case does not provide us with a significant difference with respect to the flat $\Lambda$CDM scenario.   
\begin{table}[h!]
\centering
\begin{tabular}{| c | c | c |}
\hline
Data  & \qquad Best-fit values: mean$^{+1\sigma}_{-1\sigma}$ \qquad  & Goodness of fit \\
\hline
 &  $h \qquad\qquad \Omega_{0\, m} \qquad\qquad \lambda$  & $\chi^{2}_{min}$ \quad BIC \\
\hline
\multicolumn{3}{ |c| }{CBFP} \\ 
\hline
OHD  & $0.706^{+0.015}_{-0.015} \quad 0.277^{+0.033}_{-0.043} \quad -0.019^{+0.028}_{-0.010} $ & $29.38 \quad 41.24  $ \\
\hline
Joint  & $0.704^{+0.0085}_{-0.0085} \quad 0.282^{+0.015}_{-0.015} \quad -0.018^{+0.011}_{-0.011} $ & $1055.84 \quad 1076.85  $ \\
\hline
\end{tabular}
\caption{Results of the best-fit parameters and statistical indicators
of a negative contribution to the matter sector. The uncertainties correspond to $1\sigma$ $(68.3\%)$ of the confidence level (CL). We consider the criterion $R-1<1.0\times 10^{-2}$ for the OHD data and the joint analysis.}
\label{tab:cobaya_negative}
\end{table}
%

%%%%%%%%%%%%%%%%%%%%%%%%%%%%%%%%%%%%%%%
% ACKNOWLEDGMENTS
%%%%%%%%%%%%%%%%%%%%%%%%%%%%%%%%%%%%%%%
\acknowledgments
We thank Francisco X. Linares Cede\~{n}o and Antonio Herrera-Mart\'in for useful comments and suggestions. Also, we thank Esteban Gonz\'{a}lez for helping us with the statistical analysis. L. Arturo Ureña-López gave us thorough insights about the final version of the manuscript. The authors thank the anonymous reviewer for helping us to improve our paper. This work was supported by CONACyT Network Project No. 376127 {\it Sombras, lentes y ondas gravitatorias generadas por objetos compactos astrofísicos}. R.H.J is supported by CONACYT Estancias Posdoctorales por M\'{e}xico, Modalidad 1: Estancia Posdoctoral Acad\'{e}mica. C.M. thanks PROSNI-UDG 2021 support. M. B. acknowledges CONICET, Argentina (PIP 11220200100110CO), and UNMdP (EXA955/20) for financial support.

%
%%%%%%%%%%%%%%%%%%%%%%%%%%%%%%%%
%%%%%%%  REFERENCES   %%%%%%%%%
%%%%%%%%%%%%%%%%%%%%%%%%%%%%%%%%
%
\bibliographystyle{unsrt}
\bibliography{CBFP4}

\end{document}